\documentclass[twocolumn,aps,superscriptaddress]{revtex4}
\usepackage{hyperref}
\usepackage{graphicx}
\usepackage{amsmath}
\usepackage[utf8]{inputenc}

\begin{document}

\title{Discrimination of Optical Coherent States using a Photon Number Resolving Detector}

\author{Christoffer Wittmann}
\affiliation{Max Planck Institute for the Science of Light, G\"{u}nther-Scharowsky-Str. 1, Bau 24, 91058, Erlangen, Germany}
\affiliation{Institute for Optics, Information and Photonics, University of Erlangen-Nuremberg, Staudtstra\ss e 7/B2, 91058, Erlangen, Germany}

\author{Ulrik L. Andersen}
\affiliation{Department of Physics, Technical University of Denmark, Building 309, 2800 Kgs. Lyngby, Denmark}
\affiliation{Max Planck Institute for the Science of Light, G\"{u}nther-Scharowsky-Str. 1, Bau 24, 91058, Erlangen, Germany}

\author{Gerd Leuchs}
\affiliation{Max Planck Institute for the Science of Light, G\"{u}nther-Scharowsky-Str. 1, Bau 24, 91058, Erlangen, Germany}
\affiliation{Institute for Optics, Information and Photonics, University of Erlangen-Nuremberg, Staudtstra\ss e 7/B2, 91058, Erlangen, Germany}

\date{\today}

\begin{abstract}
The discrimination of non-orthogonal quantum states with reduced or without errors is a fundamental task in quantum measurement theory. In this work, we investigate a quantum measurement strategy capable of discriminating two coherent states probabilistically with significantly smaller error probabilities than can be obtained using non-probabilistic state discrimination. We find that appropriate postselection of the measurement data of a photon number resolving detector can be used to discriminate two coherent states with small error probability. We compare our new receiver to an optimal intermediate measurement between minimum error discrimination and unambiguous state discrimination.
\end{abstract}


\maketitle
 
\section{Introduction}
Quantum mechanics puts severe limitations on our ability to determine the state of a quantum system. It is a well known fact, that the attempt to construct a measurement device, that can discriminate between non-orthogonal quantum states without error, will fail~\cite{helstrom_quantum_1976}. Suppose for example, one is given one of two a priori known coherent states $|\alpha_{\mathrm{1,2}}\rangle$ (possibly representing binary information) with the \textit{a priori} probabilities $p_{\mathrm{1,2}}$, then there is no physical measurement that with certainty can identify which state was at hand due to the intrinsic non-orthogonality of coherent states. This limits the processing of quantum information encoded in coherent states, and leads to errors in classical telecommunication. On the other hand, it allows for unconditionally secure communication via quantum key distribution~\cite{grosshans_quantum_2003}. Moreover, optimal discrimination between coherent states is an important measurement strategy in various quantum devices such as quantum computers~\cite{bergou_discrimination_2004, munro_weak_2005} and quantum repeaters~\cite{van_loock_quantum_2008}.

Since a discrimination without any ambiguity is impossible, the canonical task is to construct a measurement apparatus that maximizes the information gained or minimizes the errors in the measurement. These  different requirements lead to two approaches.
The approach maximizing the information is the minimum error state discrimination~\cite{levitin_optimal_1995}, which has been analyzed by Helstrom~\cite{helstrom_quantum_1976}. He showed, that the minimum error is obtained by a standard von
Neumann measurement which generates erroneous results with the probability $p_\mathrm{H}=\frac{1}{2}(1-(1-4p_1p_2|\langle\alpha_1|\alpha_2\rangle|^2)^{\frac{1}{2}})$. For the discrimination of two coherent states, several optimal and near-optimal receivers have been proposed~\cite{kennedy_optical_1973, dolinar__1973, sasaki_optimum_1996,  geremia_distinguishing_2004, takeoka_implementation_2005,takeoka_discrimination_2008}. Two of them were experimentally demonstrated~\cite{cook_optical_2007, wittmann_demonstration_2008}.
The second approach is the unambiguous state discrimination (USD) originally proposed by Ivanovic, Dieks and Peres~\cite{ivanovic_to_1987, dieks_overlap_1988, peres_to_1988, jaeger_optimal_1995}. The Ivanovic-Dieks-Peres measurement produces either an error-free or an inconclusive result, where the latter occurs with the probability $p_{\mathrm{IDP}}=|\langle\alpha_1|\alpha_2\rangle|$. The first scheme reaching this bound was proposed by Banaszek~\cite{banaszek_optimal_1999}. A similar receiver for a set of unknown coherent states was proposed in~\cite{sedlk_unambiguous_2007}. The receiver is based on quantum comparison of coherent states~\cite{andersson_experimentally_2006} and was recently demonstrated for two unknown coherent states~\cite{bartkov_programmable_2008}. The optimality of the receiver for unknown states is shown in~\cite{sedlk_unambiguous_2009}. Moreover, the demonstrated receiver would perform near-optimal for two {\itshape known} coherent states, which is considered in this paper. Although the experimental realizations of USD are in principal error free, the limit cannot be reached in principle due to darkcounts of single photon counting modules and other device imperfections.

Both approaches are two limiting cases of a more general (intermediate) scheme allowing for erroneous and inconclusive results. It is plausible that there exists a trade off between the rate of errors and inconclusive results, meaning that (starting from discrimination with minimum error) an increasing probability for inconclusive results can lower the error probability. In the discrimination of pure states, the minimal probability of errors for a fixed probability of inconclusive results is derived in~\cite{chefles_strategies_1998} (and for mixed states in~\cite{fiurek_optimal_2003}). 

In this paper, we investigate a new experimentally feasible receiver in the intermediate regime. The new receiver consists of an optimized displacement operation and a photon number resolving detector. The scheme is similar to the one in~\cite{wittmann_demonstration_2008}, however the photon number resolving detector replaces a simple threshold detector. We show that proper postselection of the measurement data leads to improved error rates on the expense of inconclusive results.

In the following parts of the paper, we introduce the so called intermediate measurement for coherent states in chapter~\ref{meas}, and in chapter~\ref{PNR}, we consider a receiver using an optimized displacement and a photon number resolving detector. We finally compare the results to the optimal bound in chapter~\ref{comp}.

\section{Intermediate measurement}
\label{meas}

Consider a binary alphabet of two pure and phase shifted coherent states $\{|-\alpha\rangle,|\alpha\rangle\}$ occuring with the a priori probabilities $p_1$ and $p_2$. The task of the receiver is to certify  whether the state was prepared in $|-\alpha\rangle$ or $|\alpha\rangle$ using a measurement described by the three-component positive operator-valued measure (POVM) $\hat\Pi_i, i=1,2,?$ where $\hat\Pi_i > 0$ and $\hat\Pi_1+\hat\Pi_2+\hat\Pi_?=\hat I$. The average error probability is given by 
\begin{equation}
p_\mathrm{E}=\frac{p_1\langle-\alpha|\hat\Pi_2|-\alpha\rangle+p_2\langle \alpha|\hat\Pi_1|\alpha\rangle}{1-p_\mathrm{inc}},
\label{errorrate}
\end{equation}
where $\langle-\alpha|\hat\Pi_2|-\alpha\rangle$ ($\langle \alpha|\hat\Pi_1|\alpha\rangle$) represents the conditional error probability of mistakenly guessing $|\alpha\rangle$ $(|-\alpha\rangle)$ when $|-\alpha\rangle$ $(|\alpha\rangle)$ was prepared. The probability of inconclusive results $p_\mathrm{inc}$ is given by 
\begin{equation}
p_\mathrm{inc}=p_1\langle-\alpha|\hat\Pi_?|-\alpha\rangle+p_2\langle \alpha|\hat\Pi_?|\alpha\rangle.
\label{inconclusive}
\end{equation}
where $\langle-\alpha|\hat\Pi_?|-\alpha\rangle$($\langle \alpha|\hat\Pi_?|\alpha\rangle$) represents the conditional probability for an inconclusive result when $|-\alpha\rangle$ $(|\alpha\rangle)$ was prepared. We assume $\alpha$ is real and the two states to be prepared with the same probabilities: $p_1=p_2=1/2$. (In classical communication this popular encoding is called Binary Phase Shift Keying (BPSK).)

In the case where $p_\mathrm{inc}=0$, the states are discriminated with minimum error, whereas for  $p_E=0$, the states are discriminated unambiguously. The minimum error rate in the discrimination of two pure qubit states assuming a fixed probability for inconclusive results $p_\mathrm{inc}$ is given in Ref.~\cite{chefles_strategies_1998}. After adapting it to the coherent state alphabet, it reads
\begin{equation}
p_\mathrm{E}\ge \frac{1-p_\mathrm{inc}-\left(1-2p_\mathrm{inc}(1-\sigma)-\sigma^2\right)^{1/2}}{2(1-p_\mathrm{inc})},
\label{opterror}
\end{equation}
where $\sigma=|\langle-\alpha|\alpha\rangle|$.

Note that without inconclusive results, i.e. $p_\mathrm{inc}=0$, one finds that the error probability is limited by the Helstrom bound $p_\mathrm{E}\ge p_\mathrm{H}$. Only if the inconclusive results are more probable than the Ivanovic-Dieks-Peres bound, i.e. $p_\mathrm{inc} \ge p_\mathrm{IDP}$, an error free discrimination is possible in principle. There is no proposal for an ideal measurement device that saturates  Eqn.~\ref{opterror} for the BPSK alphabet. In the following, we present a near-optimal device.

\begin{figure}
\centerline{\includegraphics[width=6cm]{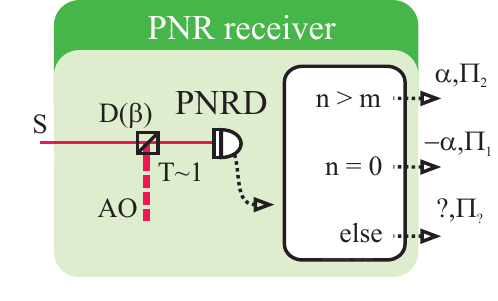}}
\caption{\label{Setup}   Schematics of the photon number resolving (PNR) receiver. The signal (S) is interfered with an auxiliary oscillator (AO) on a highly transmittive beam splitter. This results in a displacement $D(\beta)$. Subsequently, the result of a photon number resolving measurement (PNRM) is used to guess the received signal state.}
\end{figure}

\begin{figure}
\begin{tabular}{l}
(a)\\[-0.4cm]
\centerline{\includegraphics[width=8.4cm]{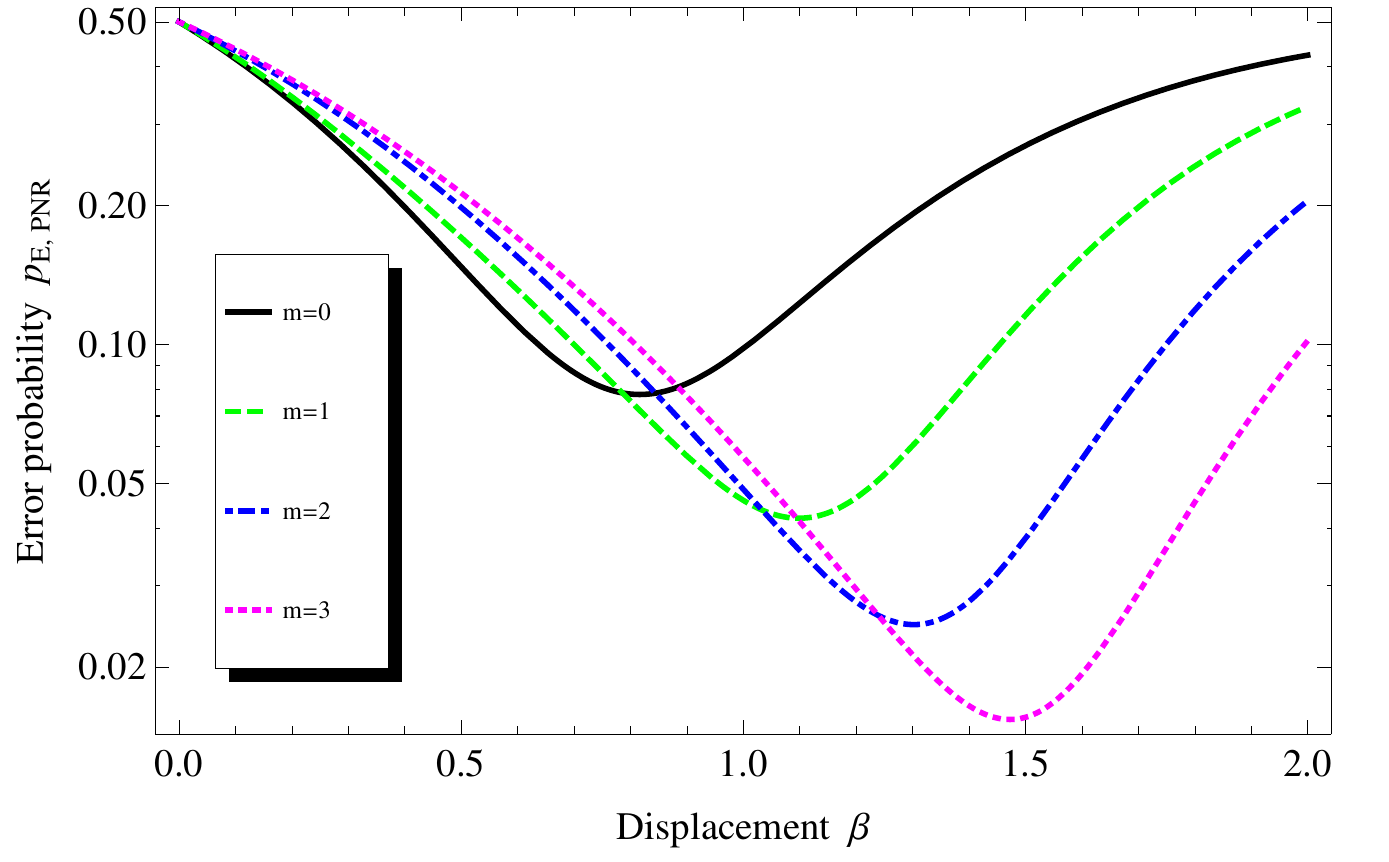}} \\ [0.3cm]
(b)\\ [-0.4cm]
\centerline{\includegraphics[width=8.4cm]{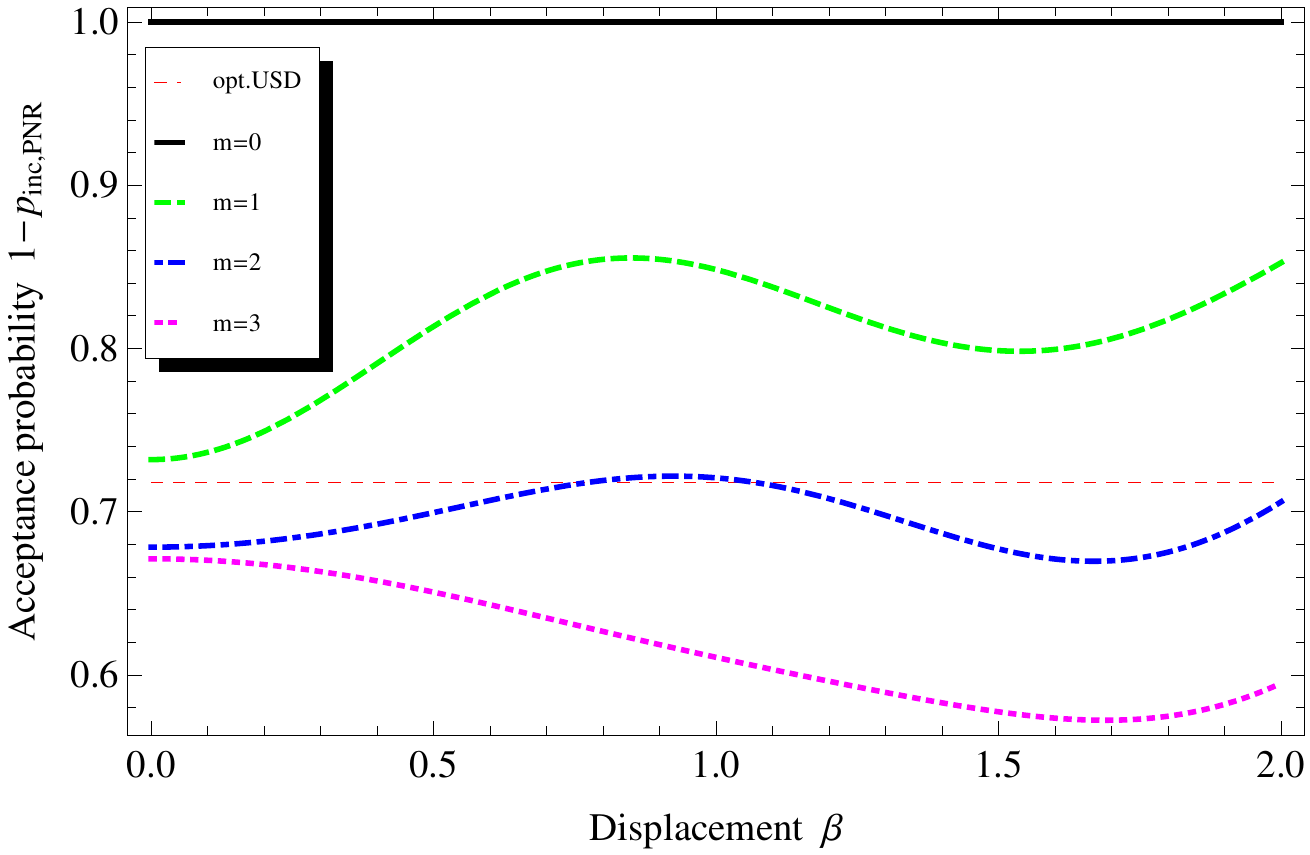}} \\

\end{tabular}
\caption{\label{displacement} (a) Effect of displacement $\beta$ on error rate for a given signal amplitude $|\alpha|^2=0.4$. Error rates for the receiver without inconclusive result (black), and the receivers with $m=1,2$ and $3$ (green,blue,purple) are shown. (b) Acceptance rates $1-p_\mathrm{inc,PNR}$ for the detection schemes with different number of dropped results $m$ for a given signal amplitude $|\alpha|^2=0.4$ (solid lines). The acceptance probability of a USD device for equal signal amplitude is also shown (thin dashed line).}
\end{figure}

\begin{figure}
\begin{tabular}{l}
(a)\\[-0.4cm]
\centerline{\includegraphics[width=8.4cm]{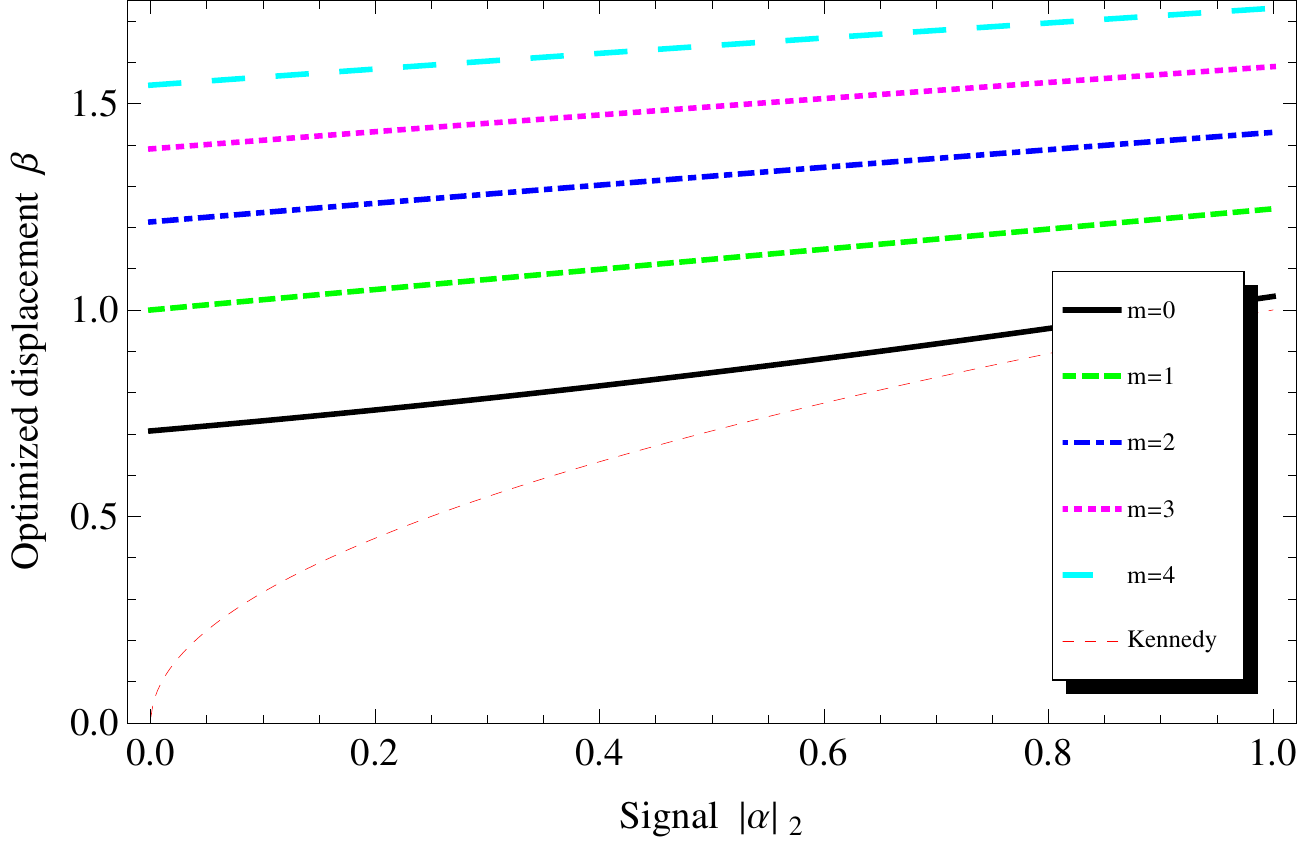}} \\ [0.3cm]
(b)\\ [-0.4cm]
\centerline{\includegraphics[width=8.4cm]{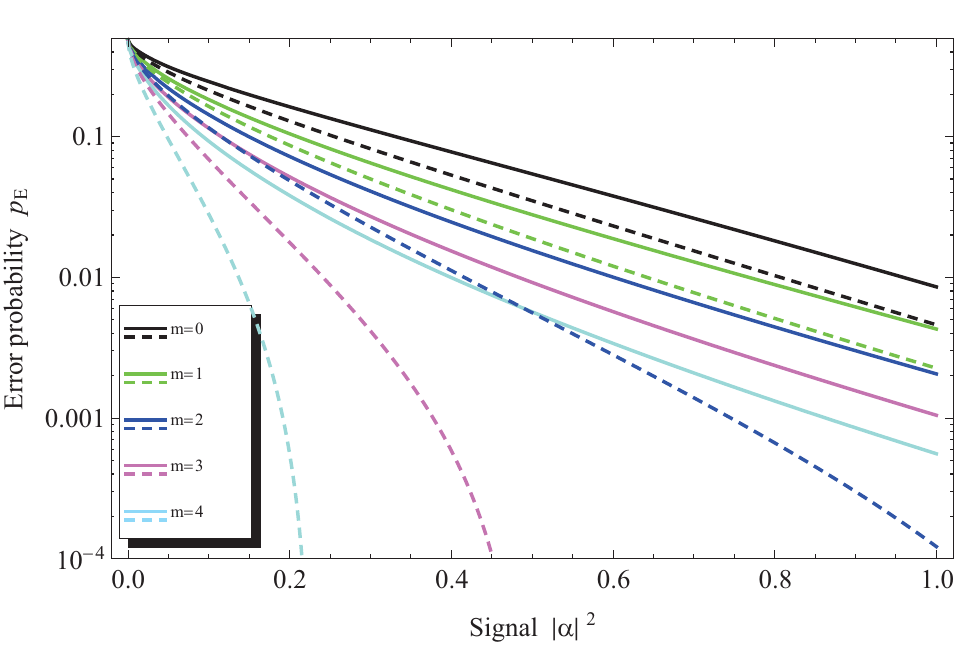}} 

\end{tabular}
\caption{\label{opt_and_error} (a) Displacement $\beta$ for a given signal amplitude $|\alpha|^2$ for different receivers: Kennedy receiver, i.e. $\beta=\alpha$ (thin red), and optimized displacement for the receivers with $m=0$ to $4$. (b) Error rates for the detection schemes ignoring different number of results $m$ (solid lines) compared to optimal detectors with equal probability for inconclusive results (dashed lines).}
\end{figure}

\begin{figure}
\begin{tabular}{l}
(a)\\[-0.4cm]
\centerline{\includegraphics[width=8.4cm]{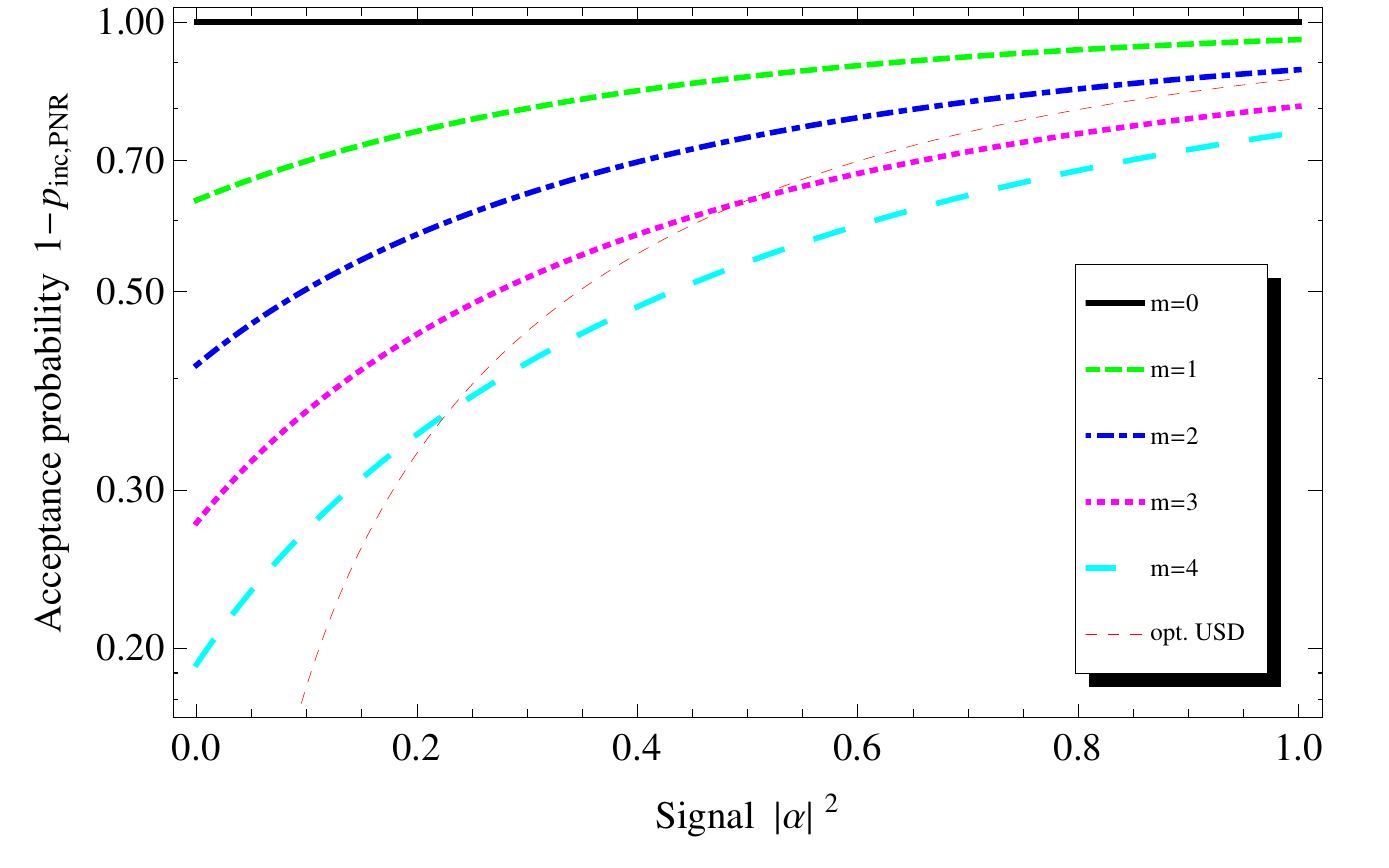}} \\ [0.3cm]
(b)\\ [-0.4cm]
\centerline{\includegraphics[width=8.4cm]{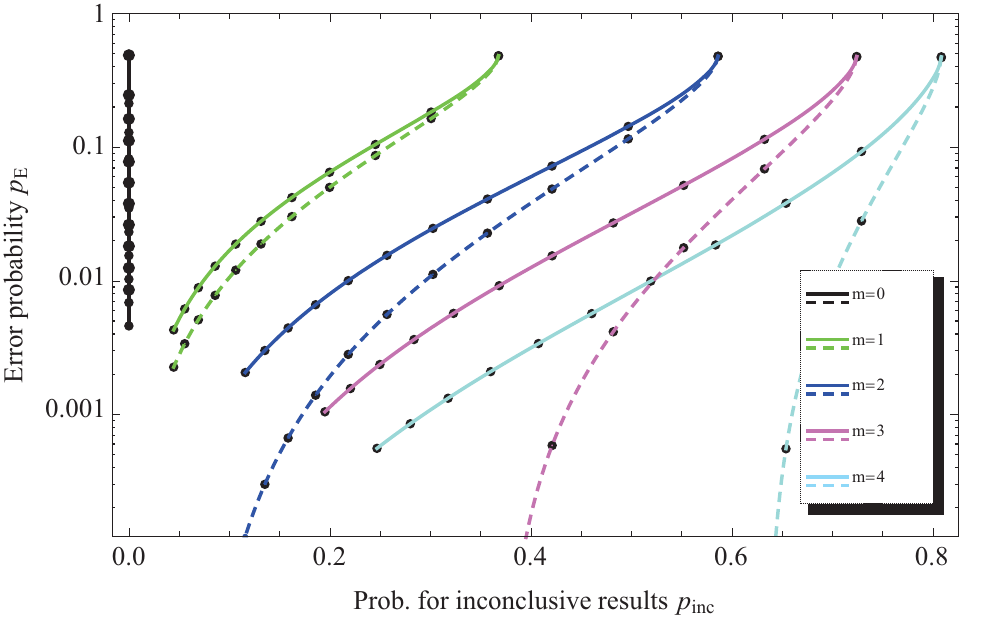}} 

\end{tabular}
\caption{\label{acc_and_comp}(a) Acceptance rate for a given signal amplitude. PNR receivers (solid lines) are compared to a perfect USD device (dashed line). (b) Dependence of error probability on the probability of inconclusive results for PNR receivers and an ideal intermediate measurement. In the parametric plot the mean photon number $\alpha|^2$ is varied from 0.002 to 1. Dots are added with a spacing of 0.1.}
\end{figure}

\section{photon number resolving receiver}
\label{PNR}

We propose a receiver, which consists of two stages: a displacement operation $D(\beta)$, where $\beta$ is real in case we also assume that $\alpha$ is real and a photon number resolving (PNR) detector. It is sketched in Fig.~\ref{Setup}.

The displacement is typically realized by a highly transmissive beamsplitter (BS) on which the signal is interfered with a much stronger auxiliary oscillator. The quality of the displacement is highly depending on the mode overlap and the BS transmittivity, otherwise straylight and loss would dominate and degrade the error rate of the receiver. Subsequently the light is detected with a PNR detector. 
In contrast to the basic on/off detection, the PNR detector allows for postselection in the photon number basis, similar to postselection of results from quadrature measurements, as e.g. in several protocols of continuous-variable quantum cryptography using homodyne detection~\cite{silberhorn_continuous_2002, lorenz_continuous-variable_2004, lance_no-switching_2005,heid_efficiency_2006}. We assume inconclusive results to occur when a small but non-zero photon number is detected. This can be described with the projector $\hat\Pi_?=\sum_{n=1}^{m}{|n\rangle\langle n|}$ (with $m>0$), where $m$ is our new postselection parameter. Conclusive results are described by $\hat\Pi_1=|0\rangle\langle 0|$ and $\hat\Pi_2=\hat I -\hat\Pi_1-\hat\Pi_?$, where $\hat\Pi_1$ identifies $|-\alpha\rangle$ and $\hat\Pi_2$ identifies $|\alpha\rangle$. The error rate is then given by (using Eqn.~\ref{errorrate})

\begin{eqnarray}
\label{pbetacheck}
p_{\mathrm{E,PNR}}=\frac{(1+2e^{-(|\alpha|^2+|\beta|^2)}\mathrm{sinh}(2\alpha\beta)}{2(1-p_{\mathrm{inc,PNR}})} \\\nonumber
-\frac{\sum_{n=1}^m \frac{|\beta-\alpha|^{2n}}{n!}e^{-|\beta-\alpha|^2}}{2(1-p_{\mathrm{inc,PNR}})} \\\nonumber
=\frac{\left(1-\frac{\Gamma \left(\text{m}+1,(\alpha -\beta )^2\right)}{\Gamma (\text{m}+1)}+ e^{-(\alpha +\beta )^2}\right)}{2(1-p_{\mathrm{inc,PNR}})},
\end{eqnarray}

where the Euler gamma funtion $\Gamma(z)$ and the incomplete gamma function $\Gamma(a,z)$ are defined as $\Gamma (z)=\int _0^{\infty }t^{z-1}e^{-t}dt$ and $\Gamma (a,z)=\int _z^{\infty }t^{a-1}e^{-t}dt$. The probability of inconclusive results is given by

\begin{eqnarray}
\label{pbetainccheck}
p_{\mathrm{inc,PNR}}=\frac{ \Gamma \left(m+1,(\alpha -\beta )^2\right)+ \Gamma \left(m+1,(\alpha +\beta )^2\right)}{2\Gamma (m+1)} \\\nonumber
- \frac{1}{2} e^{-(\alpha -\beta )^2}-\frac{1}{2} e^{-(\alpha +\beta )^2}.
\end{eqnarray}

In Fig.~\ref{displacement}(a), the dependence of the error probability $p_{\mathrm{E,PNR}}$ on the displacement $\beta$ and the postselection parameter $m$ is shown for the signal amplitude $|\alpha|^2=0.4$ (as given by Eqn.~\ref{pbetacheck}). The amplitude was chosen such that the states have a significant overlap, thereby showing strong quantum properties. The corresponding probabilities of inconclusive results are given in Eqn.~\ref{pbetainccheck} and plotted for varying displacements in Fig.~\ref{displacement}(b).

From these plots it is evident that the parameters $m$ and $\beta$ control the error probability and the corresponding probability of inconclusive results. Lower error probabilities are obtained as $m$ is increased and $\beta$ is chosen properly. Therefore, to minimize the error rate of a PNR receiver with a given $m$, the displacement $\beta$ must be optimized. Such optimization corresponds to $\mathrm{d}p_{\mathrm{E,PNR}}/\mathrm{d}\beta = 0$.
The optimal displacements for different selections of $m$ is shown in Fig.~\ref{opt_and_error}(a). When it is applied, the PNR receiver yields error probabilities shown in Fig.~\ref{opt_and_error}(b)(solid lines).

\section{Comparison of measurement strategies}
\label{comp}

We addressed the discrimination of two non-orthogonal coherent states, partly in the regime where all results are accepted but with errors, and partly in the intermediate measurement regime. Our receiver cannot reach the regime where states are discriminated without errors. It would however be very interesting (from an experimentalists point of view), to see whether a realistic experimental realisation of our strategy may outperform a practical "error-free" receiver as implemented in ref~\cite{bartkov_programmable_2008} (or the ideal proposed in Ref.~\cite{banaszek_optimal_1999}), since the measurements of such a device will also not result in error free data~\cite{bartkov_programmable_2008}. The probability of inconclusive results would then be fixed to the one achieved by the USD measurement and the task would be to reach error rates as small as possible. From a practical point of view, the PNR receiver has the advantage of being simple compared to the other implementations.

The acceptance probabilities of different PNR receivers are shown in Fig.~\ref{acc_and_comp}(a). We find that their acceptance probabilities (solid lines) are crossing the acceptance probability of a USD device and exceed them for a range of signal amplitudes. This means that our detector can be compared to an error-free device at the intersections. Since the error-free device only yields acceptance probabilities below the dashed line, whenever the PNR receivers have higher acceptance rates, they should be compared to the optimal bound given by Eqn.~\ref{opterror}. 

For the comparison of the error rates, we assume that equal signal amplitudes are sent to the PNR receiver and the optimal receiver. We choose their probabilities of inconclusive results to be equal. The error rates for our receiver (solid lines) and the optimal intermediate measurement (dashed line) are plotted in Fig.~\ref{opt_and_error}(b) for varying signal amplitudes. Alternatively, the error probabilities and rates of inconclusive results can be shown as parametric plot as illustrated in Fig.~\ref{acc_and_comp}(b). Although our receiver is not approaching the optimal limit, the error rate is significantly improved by the inconclusive results. For moderate number of dropped results in the postselection step up to $m=4$, we find a decrease of the error probability by more than one order of magnitude for signal amplitudes $|\alpha|^2>0.3$.

\section{Conclusion}

In this paper, we have addressed the discrimination of a BPSK signal involving inconlusive results in the intermediate region of USD and minimum error state discrimination. We proposed a feasible receiver consisting of an optimized displacement operation and a photon number resolving detector. The receiver drops low but non-zero photon number measurement results and therefore has an improved error rate on the expense of inconclusive results. We derived the probability of errors and the corresponding probability of inconclusive results and compared these rates with the corresponding optimal rates for USD and intermediate measurements. Although not optimal, we believe that the PNR resolving detection strategy is promising for coherent state discrimination due to its expected high efficiency in a foreseeable future. The development of single photon counting detectors concerning quantum efficiency and speed has accelerated very much in the last years and high efficiency, high speed detectors might be available in a near future. We are currently implementing the PNR receiver in our laboratory and we plan to compare its performance to that of a standard homodyne receiver.

This work has been supported by the EU project COMPAS (no. 212008) and Lundbeckfonden (no. R13-A1274). The authors would like to thank Masahiro Takeoka, Dominique Elser, Christoph Marquardt and Radim Filip for valuable discussions.


\end{document}